\documentclass[sigconf]{acmart}

\AtBeginDocument{%
  \providecommand\BibTeX{{%
    \normalfont B\kern-0.5em{\scshape i\kern-0.25em b}\kern-0.8em\TeX}}}

\copyrightyear{2024}
\acmYear{2024}
\setcopyright{rightsretained}
\acmConference[IDC '24]{Interaction Design and Children}{June 17--20, 2024}{Delft, Netherlands}
\acmBooktitle{Interaction Design and Children (IDC '24), June 17--20, 2024, Delft, Netherlands}\acmDOI{10.1145/3628516.3659392}
\acmISBN{979-8-4007-0442-0/24/06}

\usepackage{subcaption}
\usepackage{xcolor}

\begin{document}

\title[Utilizing Generative AI for App Development]{App Planner: Utilizing Generative AI in K-12 Mobile App Development Education}


\author{David Y.J. Kim}
\email{dyjkim@mit.edu}
\author{Prerna Ravi}
\email{prernar@mit.edu}
\author{Randi Williams}
\email{randi21@mit.edu}
\affiliation{%
  \institution{Massachusetts Institute of Technology}
  \streetaddress{77 Massachusetts Ave}
  \city{Cambrdige}
  \state{MA}
  \country{USA}
  \postcode{02139}
}

\author{Daeun Yoo}
\affiliation{%
  \institution{Harvard University}
  \streetaddress{Massachusetts Hall}
  \city{Cambrdige}
  \state{MA}
  \country{USA}
  \postcode{02138}
  }
\email{daeun_yoo@mde.harvard.edu}

\renewcommand{\shortauthors}{Kim, et al.}

\begin{abstract}
  App Planner is an interactive support tool for K-12 students, designed to assist in creating mobile applications. 
  By utilizing generative AI, App Planner helps students articulate the problem and solution through guided conversations via a chat-based interface. 
  It assists them in brainstorming and formulating new ideas for applications, provides feedback on those ideas, and stimulates creative thinking. 
  Here we report usability tests from our preliminary study with high-school students who appreciated App Planner for aiding the app design process and providing new viewpoints on human aspects especially the potential negative impact of their creation. 
\end{abstract}





\maketitle

\section{Introduction \& Background}

Technology's transformative power reshapes society, urging a shift from passive consumption to active participation in technological advancements. 
The Computational Action concept \cite{tissenbaum2019computational,pang2022effect} proposes nurturing individuals as proactive developers, leveraging technology to enhance societal and personal realms through hands-on engagement. 
At the heart of this transformation is computational thinking, a skill set encompassing ideation for problem-solving, and design thinking based on understanding human behavior~\cite{10.1145/1118178.1118215}. 
However, a significant gap in STEM education, particularly at the K-12 level, is the insufficient emphasis on these skills and the ethical aspects of technology, including its societal impact and the responsibility entailed in creating purposeful artifacts \cite{10.1145/3328778.3366825}.

Addressing this gap, we advocate for generative AI tools that guide individuals through the iterative design thinking process and encourage ethical and societal considerations in technology development. 
Previous research has highlighted the benefits of AI tutors and robotic mediators in enhancing student learning and fostering creativity, with intelligent agents proving instrumental in promoting self-efficacy and reflective practices \cite{mitnik2008autonomous, ali2021social, gordon2015can, park2017telling, jung2014evolution, kahn2016human}. 
These agents facilitate a reflective approach that not only improves creative processes but also the application of conceptual knowledge.
Building on this foundation, automatic scaffolding tools, including those leveraging generative AI, have been developed to offer comprehensive support in technical, metacognitive, and reflective domains \cite{williams2022constructionism}. 

\begin{figure}[t]
\centering
\includegraphics[width=\columnwidth]{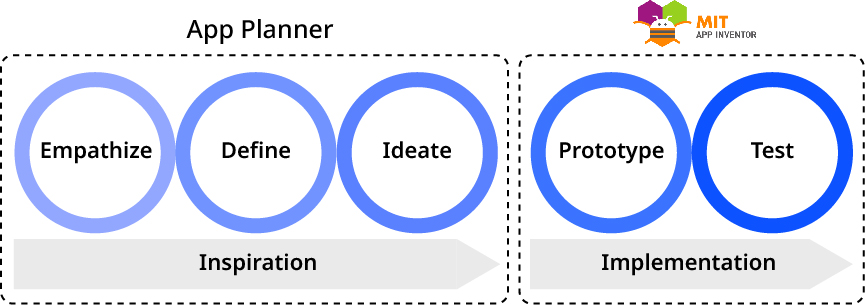}
\caption{App Planner's role in the creation process
}
\label{design}
\end{figure}

\begin{figure*}[th]
\centering
\includegraphics[width=\textwidth]{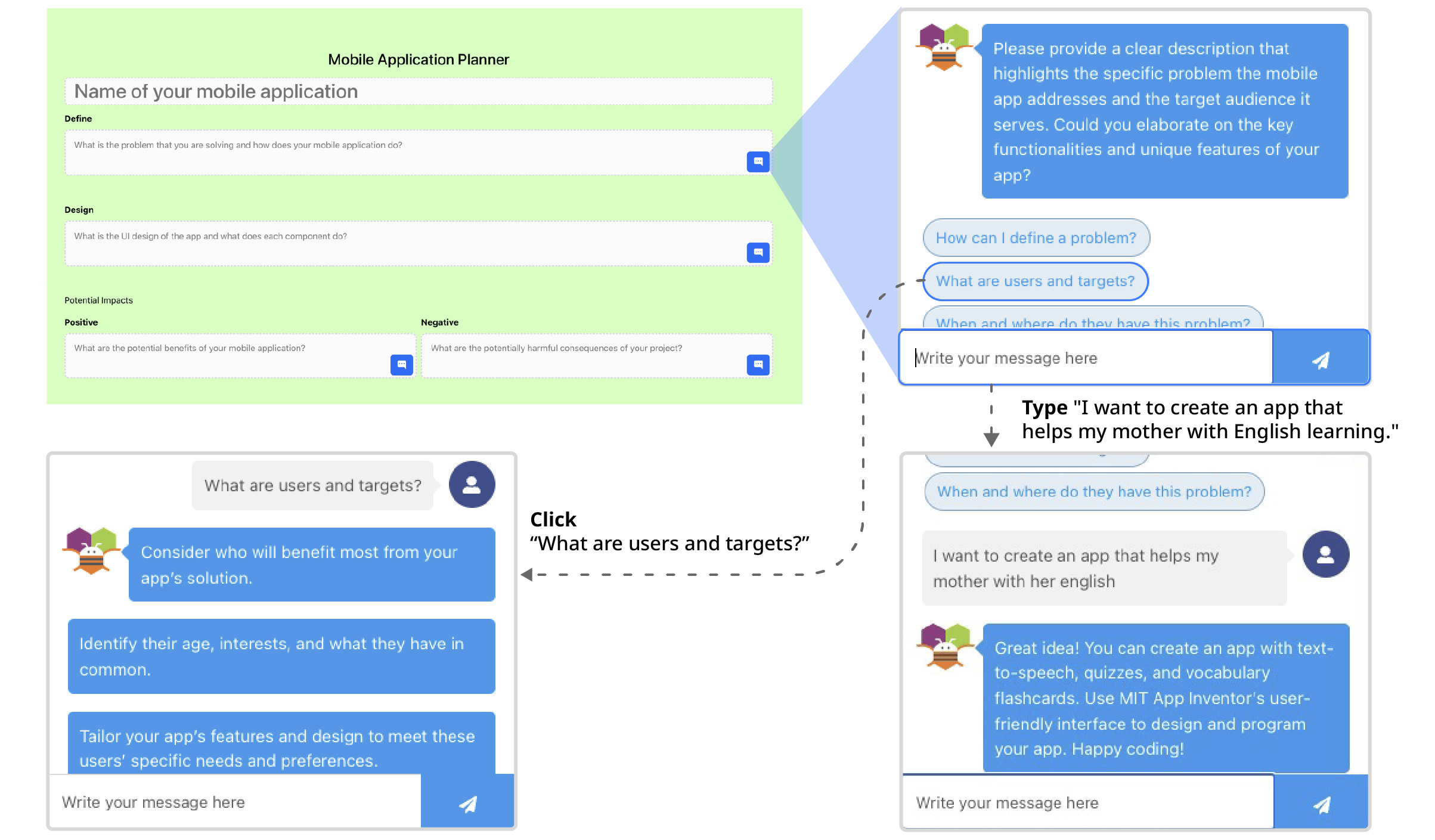}
\caption{
The tool is structured around five key text boxes, each guiding the user through a distinct stage of app development.
If students encounter difficulties while filling out any of the text boxes, they have the option to consult with a chatbot, as illustrated in the top right image. 
For immediate guidance, they can select one of the pre-set question bubbles to receive a rule-based response like the bottom left. 
Alternatively, like the bottom right students can directly input their queries, which will be processed through an API call to ChatGPT, providing them with tailored assistance and clarifications, enhancing their understanding and progress. 
\href{https://www.youtube.com/watch?v=y1pxWsf2iWE}{Video walk-through}
}
\label{workflow}
\end{figure*}

Our focus will be on mobile app development education for K-12 students~\cite{perdikuri2014students,mir2020introduction,georgiev2019students,kimadvancing}. 
For instance, a predominant tool in this educational pursuit is the MIT App Inventor platform~\cite{patton2019app,ZHOU2023LEA}. 
This platform stands out as a user-friendly visual programming environment, enabling users of all ages to craft unique applications for smartphones and tablets.
However, the platform does not accompany any tool that helps the student through the ideation of creating a mobile application.
App Planner, our proposed tool, provides a conversational interface that aids students in ideation and ethical considerations, thereby enriching the app development curriculum and fostering a comprehensive understanding of the technical and societal dimensions of app creation.
App Planner integrates generative AI to amplify the educational experience, facilitating not just technical skill development but also promoting a deep, ethical understanding of technology's impact. 
It provides continuous feedback, just-in-time support, and formative assessment, refining the app development journey while promoting metacognitive skills through divergent thinking and positive reinforcement.
Essentially, we want App Planner to inspire students in what and why they should create, along with how they should design their mobile application before jumping into the implementation process as specified in figure~\ref{design}.
This multifaceted approach demonstrates the transformative potential of AI in education, positioning the App Planner as a comprehensive platform that not only facilitates app design but also imparts critical thinking, problem-solving, and ethical reasoning, preparing students for the broader challenges of technological innovation.

\section{Tool Overview} 

We designed App Planner as a creative companion that provides technical and creative scaffolding as students work on open-ended mobile app projects. 
The overall design was inspired by S.P.A.R.K.I. (Students' Personal Assistant for Reinforcing Knowledge and Innovation)~\cite{williams2024impact}, a tool from the MIT Media lab.
It employs Open AI's GPT model~\cite{achiam2023gpt} to assess user input against a specific rubric and provide answers to users' queries. 
Additionally, it exemplifies creative thinking and delivers positive feedback. 

\subsection{Tool Design}

\begin{table*}[h]
  \caption{Paritipants}
  \label{tab:participants}
  \begin{tabular}{cclll}
    \toprule
    Participant&Age&Prior coding experience&Experience in building apps&Experience in Gen AI\\
    \midrule
    P1 & 17 & Very little & None & Some\\
    P2 & 17 & A lot & A lot & Some\\
    P3 & 17 & Very little & None & None\\
    P4 & 18 & Some & Some & Very little\\
    P5 & 17 & None & None & Very little\\
  \bottomrule
\end{tabular}
\end{table*}

App Planner (Figure~\ref{workflow}) facilitates each phase of app development for students. 
The process starts with a text box prompting students to enter the title of their project or the desired name for their mobile application. 
The next steps are based on the key features of `Design Thinking' defined by T. Brown~\cite{Brown2010DesignTF}. The organization of the sections and questions scaffold users through the design thinking and brainstorming process.

\textbf{Define}: The next text box invites students to articulate the problem they aim to address by describing the type of app they envision creating to tackle this issue.
Students are encouraged to analyze the problem they want to solve, identify target users, and consider specific contexts in which the problem occurs. This systematic approach promotes a human-centric perspective, prompting students to think critically before delving into feature development.
If they find themselves at a deadlock or struggling to progress, they have the option to engage with an intelligent chatbot.
The chatbot's role is to offer a blend of rule-based instructions for systematic guidance and AI-powered suggestions for creative enhancement.
Students are provided with interactive options such as buttons labeled `How can I define a problem?', `Who are the target users?', and `When and where do users encounter this problem?'. 
Clicking these prompts them with examples and insights to foster a deeper understanding of these aspects. 
Additionally, students can directly input queries like \textit{``I want to help my mother with her English; what kind of app should I make?''}
Such inquiries are processed through our backend, where we utilize OpenAI's API to send these requests to the GPT model~\cite{OpenAI2023GPT4TR}. 
GPT's advanced capabilities then generate tailored responses, offering students creative and practical suggestions for their app development queries.
This conversation is the cornerstone of our tool, assisting students in refining their ideas into solid goals and objectives.

\textbf{Design}: Following this, App Planner offers a dedicated section for discussing the application design. 
Here, students can deliberate on the user interface and app functionality, considering how to best structure their app to achieve the defined goals. 
For instance, a student might want to add a translator component to help their mother learn English.
This will require the student to add a text box and a button, where the mother can type in their language and press the button to see the English-translated version.
To further the design process, the Planner provides guided prompts and examples, ensuring that students' exploration remains in line with their vision.
Students can ask what features should be added and how to integrate such features into their app design to make it user-friendly.
This phase also encompasses the `Ideate' and `Prototype' stages in the Design Thinking framework.

\textbf{Positive and Negative Impact}: In the final stage, students are encouraged to thoroughly consider both the positive and potential negative impacts of their applications. 
For this critical thinking exercise the interactive chatbot serves as a brainstorming partner to explore various perspectives and consequences~\cite{Dwyer2014AnIC}. 
This step ensures that the students' creations are not only goal-oriented but also ethically sound and socially responsible. 
The inclusion of the chatbot enriches the learning experience, making it both engaging and educational.
It underscores the cultivation of a well-rounded skill set that includes app design, ethical reasoning, and critical thinking – essential attributes for the responsible creators of tomorrow.

\begin{figure}
    \centering
    \includegraphics[width=\columnwidth]{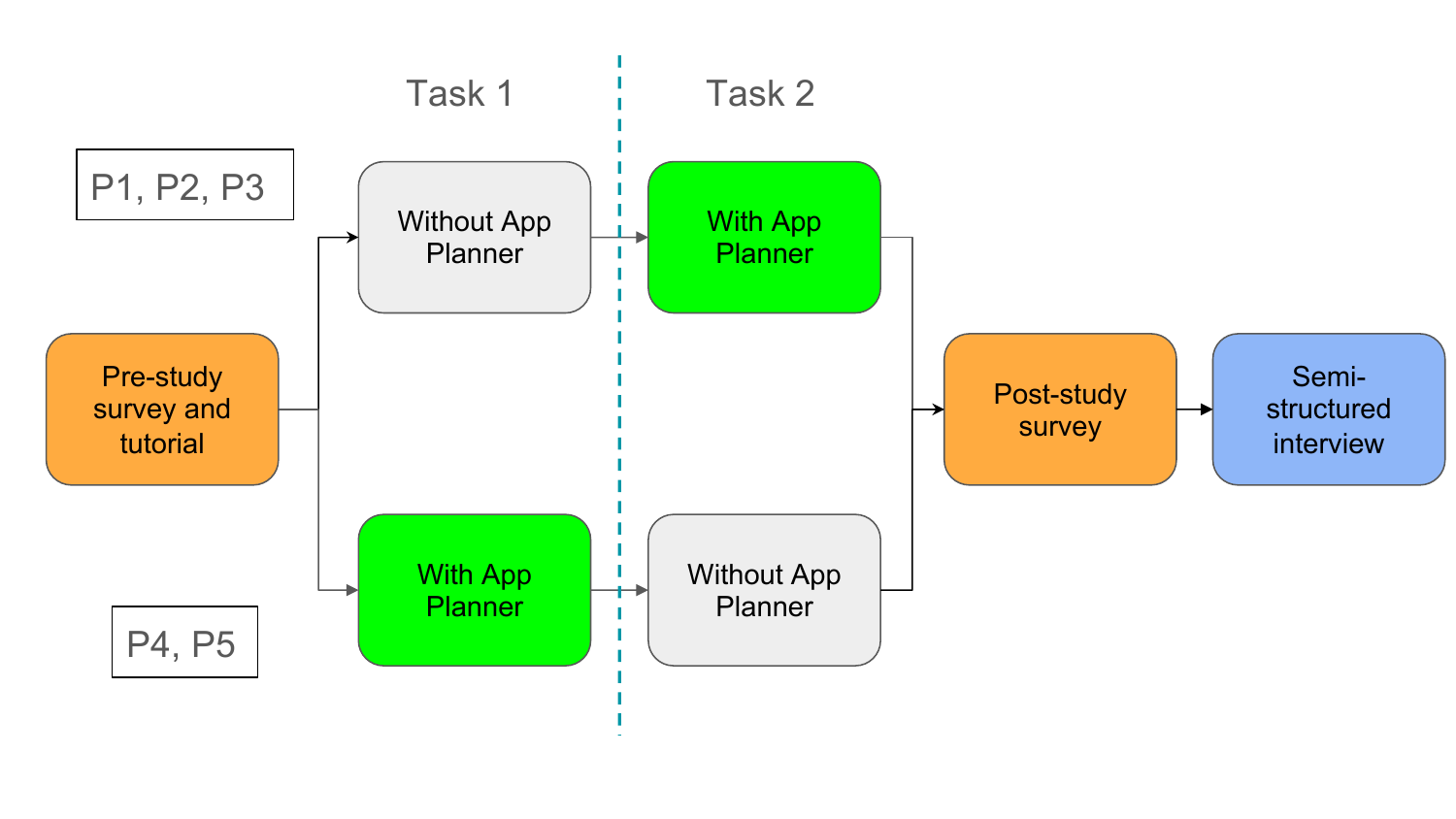}
    \caption{Study Design of the experiment}
    \label{fig:study}
\end{figure}

\section{Pilot Test}


\subsection{Participants}
Our pilot study was conducted with a cohort of five high school students (Table~\ref{tab:participants}), all in their senior year, representing a diverse range of experiences with generative AI and backgrounds in mobile application design and development. 
This varied participant group was strategically chosen to provide insights into how students with different levels of expertise and familiarity with app creation respond to the use of our tool. 
This study was conducted as part of a university class project and adhered to a rigorous Institutional Review Board (IRB) process. 
The IRB approval was obtained from the university, ensuring that all research methods, participant recruitment, and data handling procedures complied with ethical standards and regulatory guidelines.

\subsection{Methodology}
In our pilot study, we implemented a structured approach to evaluate the effectiveness of the App Planner tool. 
They were assigned two distinct design tasks, which were specifically chosen for their relevance to students' everyday experiences.
The first task required participants to conceptualize an app aiding students in day planning, with an emphasis on features like monitoring lunch menu trends to inform their meal choices. The second task focused on career exploration, asking participants to design an app that assists students in identifying potential career interests and connecting them with relevant community resources or professionals.
For each task we asked the participants to draw and describe the app they designed.

Participants were instructed to create a detailed textual description of their envisioned app for each task. To assess the impact of our App Planner tool, we varied its usage among participants. For participants P1, P2, and P3, the first task was completed without the aid of the App Planner, and they utilized the tool for the second task. Conversely, participants P4 and P5 were provided with the App Planner for the first task and completed the second task independently.
This methodology was designed to juxtapose the app design process with and without the assistance of the App Planner, thereby allowing us to analyze its efficacy in enhancing the participants' design experience and outcomes (Figure~\ref{fig:study}). 
The study's objective was to gain insights into how the tool influences the app development process, particularly in terms of creative ideation, problem-solving, and the overall usability of the tool in a real-world scenario.

Upon the completion of the two design tasks, 
we conducted semi-structured interviews with each participant. These interviews provided a platform for in-depth exploration of their experiences, allowing us to delve deeper into their thought processes, challenges encountered, and the perceived benefits or drawbacks of using the App Planner tool. The whole process took around 90 minutes for each participant.


\subsection{Observation of students using App Planner}
Every participant actively uses App Planner, but the user experience they have while interacting with App Planner has both commonalities and differences among individuals.
Overall, the stage that users prefer the most for frequent use of the AI chat feature is the \textbf{`Design' and `Negative Impact'} section. 
They use the chat feature to ask for ideas such as ``How can apply a list in my app?'' (P3), inquire about the types of resources that would help kids (P1), or seek potential output of the solution like ``What would be an example of a privacy concern for this app?'' (P4). 
They ask follow-up questions to the AI chat if the initial result from AI is obscure or too broad, such as ``What is a career exploration quiz?'' (P5).
Some participants actively use App Planner not only in the `Planning' stage but also voluntarily use chat features again in the stage of drawing the app flow and screens. 
They check what they have written in the box and chat history, or even re-ask follow-up questions in the `Design' field to gather more detailed information. 
P5 even adds a design section after drawing the app to explain the features in a more detailed way (menu), aiming to solidify ideas while incorporating them into the drawn box.
The timing for using the AI chat feature differs among individuals. 
Some participants prefer to fill out the blank by themselves initially and use the chat feature later, while others prefer to use the chat feature together at the beginning of filling out the blank. 
When we asked the participant (P5) who used the chat feature later in the semi-structured interview, she mentioned that she didn’t want to rely too much on AI, stating, ``Our opinions should be valued more when we decide what and why we need for the app, and AI can help with more technical aspects.''

\subsection{Preliminary Results \& Discussion from Semi-structured Interview}

The semi-structured interview from our pilot study also revealed many interesting viewpoints from the participants. 
The feedback from participants highlighted a strong preference for working with the App Planner as opposed to navigating the app design process independently for the following reasons.
\begin{itemize}
    \item \textbf{Easier to Proceed}: Participants found that the App Planner facilitated a more streamlined and efficient design process. 
    One participant noted, ``Using the App Planner made it easier for me to brainstorm, and I could create things much quicker. (P1),'' emphasizing the tool's ability to expedite the design phase. 
    Others mentioned how the App Planner provided crucial guidance when they felt stuck, helping them to articulate their ideas more clearly. 
    The overall agreement among participants was that, in the absence of the App Planner, the planning of subsequent steps in the app development process posed more challenges. As stated in the study (P1), ``The App Planner gave really clear instructions, so it was super easy to know what to do next.''
    \item \textbf{Practical Tips}: The participants also appreciated the practical advice offered by the App Planner. 
    P2 highlighted the value of learning specific design elements like creating a list view, which, while searchable on Google, was more conveniently accessible and contextual within the App Planner. 
    Another participant praised the creativity and helpfulness of the advice offered by the chat feature, finding it not only useful but also engaging.
    ``The chat feature in App Planner was awesome because it helped me quickly find the exact technical info I needed.'' (P5)
    \item \textbf{Expanding Thoughts}: The App Planner was also recognized for its role in broadening the participants' perspectives. 
    As they described their ideas, the tool helped them to expand upon these concepts, offering new dimensions and depth to their initial thoughts. 
    This aspect of the App Planner was crucial in enhancing the creativity and scope of the participants' app designs.
\end{itemize}


Based on responses, App Planner played a significant role in cultivating a human-centered mindset among students. 
Especially regarding the societal impact and user-centric design of their apps.

\begin{itemize}
    \item \textbf{Thinking about Users}: The App Planner also influenced how students thought about the users of their apps. 
    Without the tool, some students found themselves focused solely on their personal needs and perspectives. 
    ``When I tried making the app on my own, I noticed I wasn't thinking about different scenarios as thoroughly as I did with the App Planner.'' (P5).
    The App Planner encouraged them to consider a wider range of users as one student stated, ``Using the App Planner helped me consider the needs of a wider range of users.'' indicating a shift from a self-centric to a user-centric approach in app design.
    
    \item \textbf{Impact on Society}: All students expressed that considering the positive and negative impacts of their apps was a novel concept for them. 
    ``I usually just think about the good stuff, so it was really helpful to be asked about the possible downsides too.''  (P1), ``Thinking about the positive/negative impact of the app was new to me.'' (P5). 
    This reflects a shift in their thought process, prompted by the App Planner, to include broader social considerations in their design. 
    Another student noted the uniqueness of receiving advice related to social issues and user needs, which was not typically emphasized in their computer science classes. ``Negative \& Positive outcome was extremely helpful, I didn't learn that a lot'' (P3).
    The feature of evaluating both negative and positive outcomes was highlighted as particularly beneficial, all students mentioned that using app planner ``broadened perspectives.''
\end{itemize}


Finally, when we inquired about participants' preferred collaboration partners for app creation, they overwhelmingly chose friends, citing the creativity, shared perspectives, and enjoyment of group work. One participant noted, ``Friends can come up with creative ideas, but AI will give you common ideas'' (P2), highlighting a general consensus that while friends provide unique insights, AI tends to offer more generic solutions. 
Despite this, the App Planner was acknowledged as a valuable asset, particularly for its professional input. ``AI can provide me with more professional ideas. After establishing our own idea, we can incorporate additional insights from AI'' (P5), said one participant, indicating a preference for integrating the App Planner's structured guidance with the innovative flair of peer collaboration. 


\section{Limitation \& Future Directions}
In assessing the limitations of our pilot study involving the App Planner, several key factors emerge that could affect the robustness and applicability of our findings. First, the small sample size and its composition—exclusively high school students—limit the generalizability of our results. Recognizing this, future studies are planned to involve a broader demographic, specifically expanding the participant pool to include middle school and possibly younger students. This expansion will help us better understand how different age groups interact with the App Planner and whether the tool's effectiveness varies across educational levels.

Moreover, beyond expanding the participant demographic, a critical area of investigation is the long-term impact of using such generative AI tools in educational settings. While the App Planner has shown promising initial results in enhancing student learning and creativity, there is concern that reliance on AI-generated solutions could eventually lead to a plateau in creative exploration among students. The potential for these tools to deliver repetitive, generic answers may inadvertently stifle creativity rather than foster it. To address this, longitudinal studies are necessary to determine if the App Planner encourages sustained creative growth or if its utility diminishes over time as students become accustomed to its suggestions. This will help in refining the tool to better support continuous, creative learning without leading to dependency or creative stagnation.

In our future endeavors, we aim to strategically align our platform with established educational technology platforms, notably App Inventor~\cite{patton2019app}. 
A key focus of this integration will be collaborating with the ongoing development of `Aptly', a new research conducted by the MIT App Inventor team~\cite{kim2022speak}. 
Aptly is a cutting-edge platform designed to enable app creation through natural language inputs. 
For instance, a user could state, ``Make an app that translates English to any five languages'' and Aptly would bring this idea to life as a fully functional app.
Additionally, users can also edit their app during the creation process using natural language~\cite{granquist2023ai}.
The team believes that such a platform will democratize mobile app creation and make technology education more accessible to everyone, reaching one step closer to Computational Action.
However. the promise of making programming significantly easier might not hold entirely true.
The effectiveness of these tools depends heavily on the user's ability to first ideate the artifact and precisely articulate their requirements and intentions~\cite{kim2023redefining}.
The App Planner is poised to be instrumental in addressing these challenges by assisting users in the ideation and articulation phases of app development. 
As an intermediary, it will alleviate communication difficulties and smooth the design process in conjunction with Aptly. 
Beyond just bridging gaps, the App Planner could introduce features like interactive tutorials or context-sensitive help, enhancing users’ understanding of how to effectively convey their app concepts. 
Additionally, integrating user feedback mechanisms could further refine the tool's functionality, ensuring it evolves to meet the diverse needs of its users. 
This synergy between the App Planner and Aptly promises to enrich the app development experience, making it more intuitive and user-friendly.

\begin{acks}
We thank Kate Moore and Safinah Ali for giving valuable feedback and guidance during the whole process. We also thank Evan Patton for giving feedback during the tool creation and Hal Abelson and Randall Davis for giving valuable feedback.
\end{acks}

\bibliographystyle{ACM-Reference-Format}
\bibliography{reference}


\begin{thebibliography}{25}


\ifx \showCODEN    \undefined \def \showCODEN     #1{\unskip}     \fi
\ifx \showDOI      \undefined \def \showDOI       #1{#1}\fi
\ifx \showISBNx    \undefined \def \showISBNx     #1{\unskip}     \fi
\ifx \showISBNxiii \undefined \def \showISBNxiii  #1{\unskip}     \fi
\ifx \showISSN     \undefined \def \showISSN      #1{\unskip}     \fi
\ifx \showLCCN     \undefined \def \showLCCN      #1{\unskip}     \fi
\ifx \shownote     \undefined \def \shownote      #1{#1}          \fi
\ifx \showarticletitle \undefined \def \showarticletitle #1{#1}   \fi
\ifx \showURL      \undefined \def \showURL       {\relax}        \fi
\providecommand\bibfield[2]{#2}
\providecommand\bibinfo[2]{#2}
\providecommand\natexlab[1]{#1}
\providecommand\showeprint[2][]{arXiv:#2}

\bibitem[Achiam et~al\mbox{.}(2023)]%
        {achiam2023gpt}
\bibfield{author}{\bibinfo{person}{Josh Achiam}, \bibinfo{person}{Steven Adler}, \bibinfo{person}{Sandhini Agarwal}, \bibinfo{person}{Lama Ahmad}, \bibinfo{person}{Ilge Akkaya}, \bibinfo{person}{Florencia~Leoni Aleman}, \bibinfo{person}{Diogo Almeida}, \bibinfo{person}{Janko Altenschmidt}, \bibinfo{person}{Sam Altman}, \bibinfo{person}{Shyamal Anadkat}, {et~al\mbox{.}}} \bibinfo{year}{2023}\natexlab{}.
\newblock \showarticletitle{Gpt-4 technical report}.
\newblock \bibinfo{journal}{\emph{arXiv preprint arXiv:2303.08774}} (\bibinfo{year}{2023}).
\newblock


\bibitem[Ali et~al\mbox{.}(2021)]%
        {ali2021social}
\bibfield{author}{\bibinfo{person}{Safinah Ali}, \bibinfo{person}{Nisha Devasia}, \bibinfo{person}{Hae~Won Park}, {and} \bibinfo{person}{Cynthia Breazeal}.} \bibinfo{year}{2021}\natexlab{}.
\newblock \showarticletitle{Social robots as creativity eliciting agents}.
\newblock \bibinfo{journal}{\emph{Frontiers in Robotics and AI}}  \bibinfo{volume}{8} (\bibinfo{year}{2021}), \bibinfo{pages}{673730}.
\newblock


\bibitem[Brown and Wyatt(2010)]%
        {Brown2010DesignTF}
\bibfield{author}{\bibinfo{person}{Tim Brown} {and} \bibinfo{person}{Jocelyn Wyatt}.} \bibinfo{year}{2010}\natexlab{}.
\newblock \showarticletitle{Design Thinking for Social Innovation}.
\newblock
\urldef\tempurl%
\url{https://api.semanticscholar.org/CorpusID:6740624}
\showURL{%
\tempurl}


\bibitem[Dwyer et~al\mbox{.}(2014)]%
        {Dwyer2014AnIC}
\bibfield{author}{\bibinfo{person}{Christopher~P. Dwyer}, \bibinfo{person}{Michael~J. Hogan}, {and} \bibinfo{person}{Ian Stewart}.} \bibinfo{year}{2014}\natexlab{}.
\newblock \showarticletitle{An integrated critical thinking framework for the 21st century}.
\newblock \bibinfo{journal}{\emph{Thinking Skills and Creativity}}  \bibinfo{volume}{12} (\bibinfo{year}{2014}), \bibinfo{pages}{43--52}.
\newblock
\urldef\tempurl%
\url{https://api.semanticscholar.org/CorpusID:145570310}
\showURL{%
\tempurl}


\bibitem[Fiesler et~al\mbox{.}(2020)]%
        {10.1145/3328778.3366825}
\bibfield{author}{\bibinfo{person}{Casey Fiesler}, \bibinfo{person}{Natalie Garrett}, {and} \bibinfo{person}{Nathan Beard}.} \bibinfo{year}{2020}\natexlab{}.
\newblock \showarticletitle{What Do We Teach When We Teach Tech Ethics? A Syllabi Analysis}. In \bibinfo{booktitle}{\emph{Proceedings of the 51st ACM Technical Symposium on Computer Science Education}} (Portland, OR, USA) \emph{(\bibinfo{series}{SIGCSE '20})}. \bibinfo{publisher}{Association for Computing Machinery}, \bibinfo{address}{New York, NY, USA}, \bibinfo{pages}{289–295}.
\newblock
\showISBNx{9781450367936}
\urldef\tempurl%
\url{https://doi.org/10.1145/3328778.3366825}
\showDOI{\tempurl}


\bibitem[Georgiev(2019)]%
        {georgiev2019students}
\bibfield{author}{\bibinfo{person}{Ts~St Georgiev}.} \bibinfo{year}{2019}\natexlab{}.
\newblock \showarticletitle{Students’ viewpoint about using MIT app inventor in education}. In \bibinfo{booktitle}{\emph{2019 42nd International Convention on Information and Communication Technology, Electronics and Microelectronics (MIPRO)}}. IEEE, \bibinfo{pages}{611--616}.
\newblock


\bibitem[Gordon et~al\mbox{.}(2015)]%
        {gordon2015can}
\bibfield{author}{\bibinfo{person}{Goren Gordon}, \bibinfo{person}{Cynthia Breazeal}, {and} \bibinfo{person}{Susan Engel}.} \bibinfo{year}{2015}\natexlab{}.
\newblock \showarticletitle{Can children catch curiosity from a social robot?}. In \bibinfo{booktitle}{\emph{Proceedings of the tenth annual ACM/IEEE international conference on human-robot interaction}}. \bibinfo{pages}{91--98}.
\newblock


\bibitem[Granquist et~al\mbox{.}(2023)]%
        {granquist2023ai}
\bibfield{author}{\bibinfo{person}{Ashley~M Granquist}, \bibinfo{person}{David~YJ Kim}, {and} \bibinfo{person}{Evan~W Patton}.} \bibinfo{year}{2023}\natexlab{}.
\newblock \showarticletitle{AI-Augmented Feature to Edit and Design Mobile Applications}. In \bibinfo{booktitle}{\emph{Proceedings of the 25th International Conference on Mobile Human-Computer Interaction}}. \bibinfo{pages}{1--5}.
\newblock


\bibitem[Jung(2014)]%
        {jung2014evolution}
\bibfield{author}{\bibinfo{person}{Rex~E Jung}.} \bibinfo{year}{2014}\natexlab{}.
\newblock \bibinfo{title}{Evolution, creativity, intelligence, and madness:“Here Be Dragons”}.
\newblock , \bibinfo{numpages}{784}~pages.
\newblock


\bibitem[Kahn et~al\mbox{.}(2016)]%
        {kahn2016human}
\bibfield{author}{\bibinfo{person}{Peter~H Kahn}, \bibinfo{person}{Takayuki Kanda}, \bibinfo{person}{Hiroshi Ishiguro}, \bibinfo{person}{Brian~T Gill}, \bibinfo{person}{Solace Shen}, \bibinfo{person}{Jolina~H Ruckert}, {and} \bibinfo{person}{Heather~E Gary}.} \bibinfo{year}{2016}\natexlab{}.
\newblock \showarticletitle{Human creativity can be facilitated through interacting with a social robot}. In \bibinfo{booktitle}{\emph{2016 11th ACM/IEEE International Conference on Human-Robot Interaction (HRI)}}. IEEE, \bibinfo{pages}{173--180}.
\newblock


\bibitem[Kim(2023)]%
        {kim2023redefining}
\bibfield{author}{\bibinfo{person}{David~YJ Kim}.} \bibinfo{year}{2023}\natexlab{}.
\newblock \showarticletitle{Redefining Computer Science Education: Code-Centric to Natural Language Programming with AI-Based No-Code Platforms}.
\newblock \bibinfo{journal}{\emph{arXiv preprint arXiv:2308.13539}} (\bibinfo{year}{2023}).
\newblock


\bibitem[Kim et~al\mbox{.}(2022)]%
        {kim2022speak}
\bibfield{author}{\bibinfo{person}{David~YJ Kim}, \bibinfo{person}{Ashley Granquist}, \bibinfo{person}{Evan Patton}, \bibinfo{person}{Mark Friedman}, {and} \bibinfo{person}{Hal Abelson}.} \bibinfo{year}{2022}\natexlab{}.
\newblock \showarticletitle{Speak your mind: Introducing aptly, the software platform that turns ideas into working apps}. In \bibinfo{booktitle}{\emph{ICERI2022 Proceedings}}. IATED, \bibinfo{pages}{1653--1660}.
\newblock


\bibitem[Kim et~al\mbox{.}({[n.\,d.]})]%
        {kimadvancing}
\bibfield{author}{\bibinfo{person}{David~YJ Kim}, \bibinfo{person}{Anqi Zhou}, \bibinfo{person}{Yasuhiro Sudo}, {and} \bibinfo{person}{Kosuke Takano}.} \bibinfo{year}{[n.\,d.]}\natexlab{}.
\newblock \showarticletitle{Advancing Mobile App Development and Generative AI Education through MIT App}.
\newblock  (\bibinfo{year}{[n.\,d.]}).
\newblock


\bibitem[Mir and Llueca(2020)]%
        {mir2020introduction}
\bibfield{author}{\bibinfo{person}{Sergio~Barrachina Mir} {and} \bibinfo{person}{Germ{\'a}n~Fabregat Llueca}.} \bibinfo{year}{2020}\natexlab{}.
\newblock \showarticletitle{Introduction to programming using mobile phones and MIT app inventor}.
\newblock \bibinfo{journal}{\emph{IEEE Revista Iberoamericana de Tecnolog{\'\i}as del Aprendizaje}} \bibinfo{volume}{15}, \bibinfo{number}{3} (\bibinfo{year}{2020}), \bibinfo{pages}{192--201}.
\newblock


\bibitem[Mitnik et~al\mbox{.}(2008)]%
        {mitnik2008autonomous}
\bibfield{author}{\bibinfo{person}{Ruben Mitnik}, \bibinfo{person}{Miguel Nussbaum}, {and} \bibinfo{person}{Alvaro Soto}.} \bibinfo{year}{2008}\natexlab{}.
\newblock \showarticletitle{An autonomous educational mobile robot mediator}.
\newblock \bibinfo{journal}{\emph{Autonomous Robots}}  \bibinfo{volume}{25} (\bibinfo{year}{2008}), \bibinfo{pages}{367--382}.
\newblock


\bibitem[OpenAI(2023)]%
        {OpenAI2023GPT4TR}
\bibfield{author}{\bibinfo{person}{OpenAI}.} \bibinfo{year}{2023}\natexlab{}.
\newblock \showarticletitle{GPT-4 Technical Report}.
\newblock \bibinfo{journal}{\emph{ArXiv}}  \bibinfo{volume}{abs/2303.08774} (\bibinfo{year}{2023}).
\newblock
\urldef\tempurl%
\url{https://api.semanticscholar.org/CorpusID:257532815}
\showURL{%
\tempurl}


\bibitem[Pang et~al\mbox{.}(2022)]%
        {pang2022effect}
\bibfield{author}{\bibinfo{person}{HN Pang}, \bibinfo{person}{R Parks}, \bibinfo{person}{C Breazeal}, {and} \bibinfo{person}{H Abelson}.} \bibinfo{year}{2022}\natexlab{}.
\newblock \showarticletitle{THE EFFECT OF THE COMPUTATIONAL ACTION PROCESS ON STUDENTS'SOFTWARE PROTOTYPE IDEAS}. In \bibinfo{booktitle}{\emph{ICERI2022 Proceedings}}. IATED, \bibinfo{pages}{1804--1814}.
\newblock


\bibitem[Park et~al\mbox{.}(2017)]%
        {park2017telling}
\bibfield{author}{\bibinfo{person}{Hae~Won Park}, \bibinfo{person}{Mirko Gelsomini}, \bibinfo{person}{Jin~Joo Lee}, {and} \bibinfo{person}{Cynthia Breazeal}.} \bibinfo{year}{2017}\natexlab{}.
\newblock \showarticletitle{Telling stories to robots: The effect of backchanneling on a child's storytelling}. In \bibinfo{booktitle}{\emph{Proceedings of the 2017 ACM/IEEE international conference on human-robot interaction}}. \bibinfo{pages}{100--108}.
\newblock


\bibitem[Patton et~al\mbox{.}(2019)]%
        {patton2019app}
\bibfield{author}{\bibinfo{person}{Evan~W Patton}, \bibinfo{person}{Michael Tissenbaum}, {and} \bibinfo{person}{Farzeen Harunani}.} \bibinfo{year}{2019}\natexlab{}.
\newblock \showarticletitle{MIT app inventor: Objectives, design, and development}.
\newblock \bibinfo{journal}{\emph{Computational thinking education}} (\bibinfo{year}{2019}), \bibinfo{pages}{31--49}.
\newblock


\bibitem[Perdikuri(2014)]%
        {perdikuri2014students}
\bibfield{author}{\bibinfo{person}{Katerina Perdikuri}.} \bibinfo{year}{2014}\natexlab{}.
\newblock \showarticletitle{Students' Experiences from the use of MIT App Inventor in classroom}. In \bibinfo{booktitle}{\emph{Proceedings of the 18th Panhellenic conference on informatics}}. \bibinfo{pages}{1--6}.
\newblock


\bibitem[Tissenbaum et~al\mbox{.}(2019)]%
        {tissenbaum2019computational}
\bibfield{author}{\bibinfo{person}{Mike Tissenbaum}, \bibinfo{person}{Josh Sheldon}, {and} \bibinfo{person}{Hal Abelson}.} \bibinfo{year}{2019}\natexlab{}.
\newblock \showarticletitle{From computational thinking to computational action}.
\newblock \bibinfo{journal}{\emph{Commun. ACM}} \bibinfo{volume}{62}, \bibinfo{number}{3} (\bibinfo{year}{2019}), \bibinfo{pages}{34--36}.
\newblock


\bibitem[Williams(2022)]%
        {williams2022constructionism}
\bibfield{author}{\bibinfo{person}{Randi Williams}.} \bibinfo{year}{2022}\natexlab{}.
\newblock \showarticletitle{Constructionism, Ethics, and Creativity: Developing Tools for the Future of Education with AI}. In \bibinfo{booktitle}{\emph{2022 IEEE Symposium on Visual Languages and Human-Centric Computing (VL/HCC)}}. IEEE, \bibinfo{pages}{1--3}.
\newblock


\bibitem[Williams(2024)]%
        {williams2024impact}
\bibfield{author}{\bibinfo{person}{Randi Williams}.} \bibinfo{year}{2024}\natexlab{}.
\newblock \emph{\bibinfo{title}{Impact. AI: Democratizing AI through K-12 Artificial Intelligence Education}}.
\newblock \bibinfo{thesistype}{Ph.\,D. Dissertation}. \bibinfo{school}{Massachusetts Institute of Technology}.
\newblock


\bibitem[Wing(2006)]%
        {10.1145/1118178.1118215}
\bibfield{author}{\bibinfo{person}{Jeannette~M. Wing}.} \bibinfo{year}{2006}\natexlab{}.
\newblock \showarticletitle{Computational Thinking}.
\newblock \bibinfo{journal}{\emph{Commun. ACM}} \bibinfo{volume}{49}, \bibinfo{number}{3} (\bibinfo{date}{mar} \bibinfo{year}{2006}), \bibinfo{pages}{33–35}.
\newblock
\showISSN{0001-0782}
\urldef\tempurl%
\url{https://doi.org/10.1145/1118178.1118215}
\showDOI{\tempurl}


\bibitem[Zhou et~al\mbox{.}(2023)]%
        {ZHOU2023LEA}
\bibfield{author}{\bibinfo{person}{A. Zhou}, \bibinfo{person}{D. Kim}, {and} \bibinfo{person}{H. Abelson}.} \bibinfo{year}{2023}\natexlab{}.
\newblock \showarticletitle{LEARNING AND USING IMAGE CLASSIFIERS BY CREATING REAL MOBILE APPLICATIONS}. In \bibinfo{booktitle}{\emph{INTED2023 Proceedings}} (Valencia, Spain) \emph{(\bibinfo{series}{17th International Technology, Education and Development Conference})}. \bibinfo{publisher}{IATED}, \bibinfo{pages}{3834--3841}.
\newblock
\showISBNx{978-84-09-49026-4}
\showISSN{2340-1079}
\urldef\tempurl%
\url{https://doi.org/10.21125/inted.2023.1023}
\showDOI{\tempurl}


\end{thebibliography}










\end{document}